\documentclass[twocolumn,showpacs,showkeys,preprintnumbers,amsmath,amssymb,prl]{revtex4}

\usepackage{graphics,color}

\usepackage{graphicx}
\DeclareGraphicsExtensions{.eps,.pdf}
\usepackage{epstopdf}
\begin{document}

\title{Weyl Fermions in antiferromagnetic Mn$_3$Sn and Mn$_3$Ge}
\author{J\"urgen K\"ubler$^1$}
 \email{juergen.kuebler@gmail.com}
\author{Claudia Felser$^2$}
\affiliation{%
 $^1$Technische Universit{\"a}t Darmstadt, Germany }%
\affiliation{%
 $^2$Max Planck Institute for Chemical Physics of Solids, Dresden, Germany }%
\date{\today}
\begin{abstract}{The anomalous Hall effect in the noncollinear antiferromagnetic metals Mn$_3$Ge and Mn$_3$Sn has been observed after a theoretical prediction made by us (EPL, \textbf{108}, (2014), 67001). The experimental values of the anomalous Hall conductivities (AHC) are large as are the theoretical values. Recently measured thermoelectric properties mirror the large AHC and clearly show that the transport is by quasiparticles at the Fermi energy.    We here make an attempt to unravel the origin of the large AHC and show that both Mn$_3$Sn and Mn$_3$Ge host Weyl points, which were recently discovered in semimetals. For this purpose we determine the electronic structure \textit{ab initio} in the local spin-density functional approximation. The Weyl points are found to occur below the Fermi energy and we argue that spots of large Berry flux ('hot spots') that are seen at the Fermi surface are produced by the Weyl nodes. }   
\end{abstract}
\pacs{75.50.Ee, 75.47.Np, 73.22.Gk, 75.70.Tj}
\keywords{Weyl in antiferromagnets}

\maketitle
\textbf{Introduction}-
Recently we predicted the occurrence of an anomalous Hall effect (AHE) in the antiferromagnetic compounds Mn$_3$Sn and Mn$_3$Ge \cite{KandF} and, indeed, this effect was subsequently discovered in Mn$_3$Sn by Nakatsuji \textit{et al.} \cite{nakatsuji} as well as in Mn$_3$Ge by Nayak \textit{et al.} \cite{ajaya} and independently by Kiyohara and Nakatsuji.\cite{naoki}

While the AHE commonly occurs in ferromagnets \cite{nagaosa}, it is expected to be absent in antiferromagnets, but Chen \textit{et al.} \cite{chen} found that under certain symmetry conditions it is observable in non-collinear antiferromagnetic compounds. Extending the class of compounds possessing an AHE is of fundamental interest, but it might also be of importance for spintronics applications.\cite{baltz} 

\begin{figure}[h]
\begin{center}
 \includegraphics[width=1.0\linewidth]{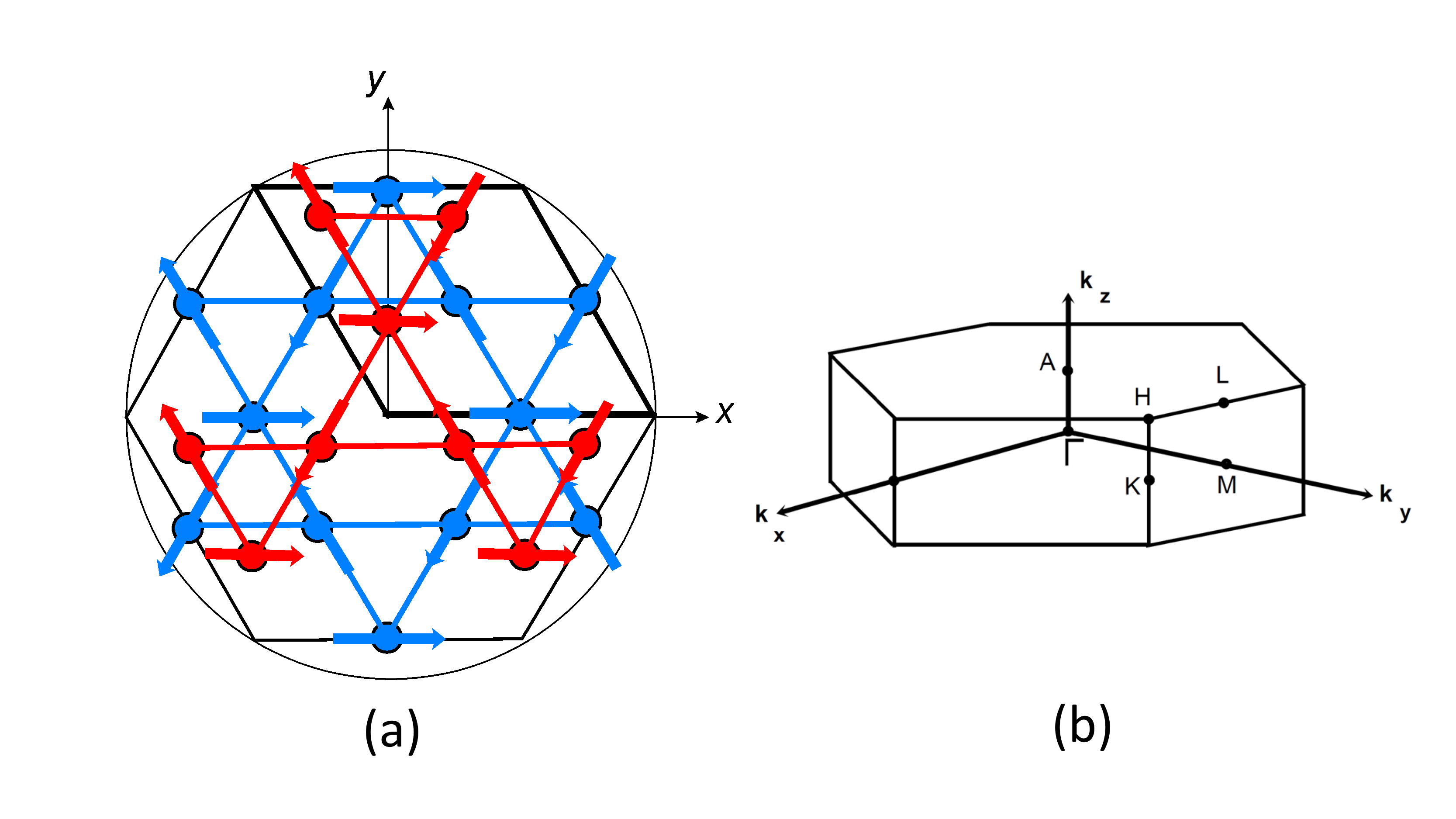}
\caption{\label{fig1} The case of Mn$_3$Sn. (a) Top view of the crystal structure indicating the directions of the magnetic moments by arrows, blue in the $z=0$ plane and red in the $z=1/2$ plane. The unit cell is enclosed by heavy solid lines, the corners of which are defined by the Sn sites in the $z=0$ plane. (b) Hexagonal Brillouin Zone. }
\end{center}
\end{figure}
\begin{figure}[h]
\begin{center}
 \includegraphics[width=1.0\linewidth]{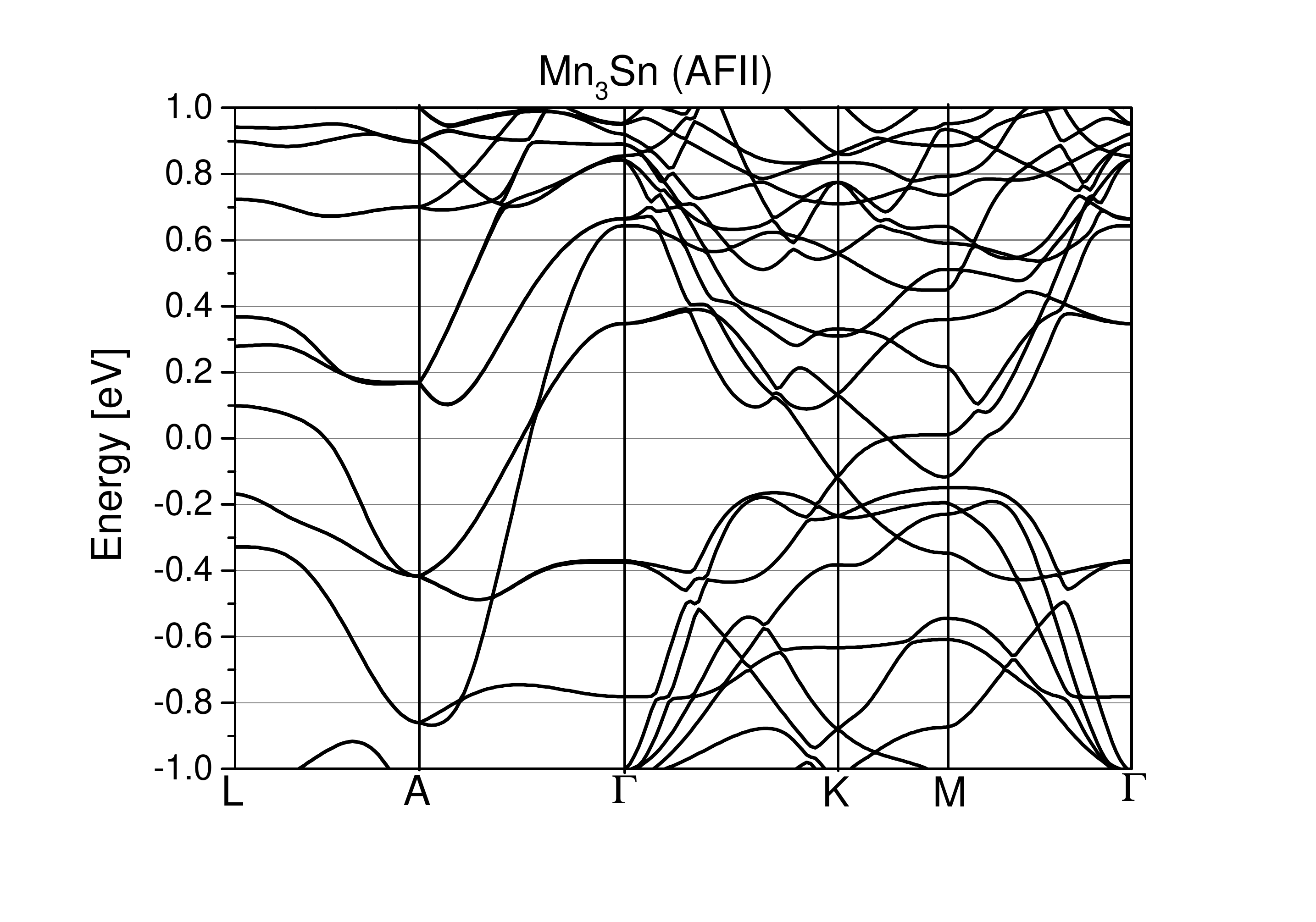}
\caption{\label{fig2}The band structure of Mn$_3$Sn plotted along  symmetry lines. The labels are defined in Fig.\ref{fig1}(b). The Fermi energy is at 0 eV.  }
\end{center}
\end{figure}
The compounds Mn$_3$Sn and Mn$_3$Ge can be grown in the hexagonal crystal structure having the symmetry P6$_3/mmc$ and a non-symmorphic spüace group. They have been shown to possess a variety of non-collinear antiferromagnetic orders, \cite{kadar71,nagamiya,tomi,sandr96,zhang} of which one  kind is sketched in Fig.~\ref{fig1}a, another one in Fig.~\ref{fig5}a. Of all the possibilities these two cases have \textit{negative chirality} \cite{sandr96}.
\begin{figure}[h]
\begin{center}
 \includegraphics[width=1.05\linewidth]{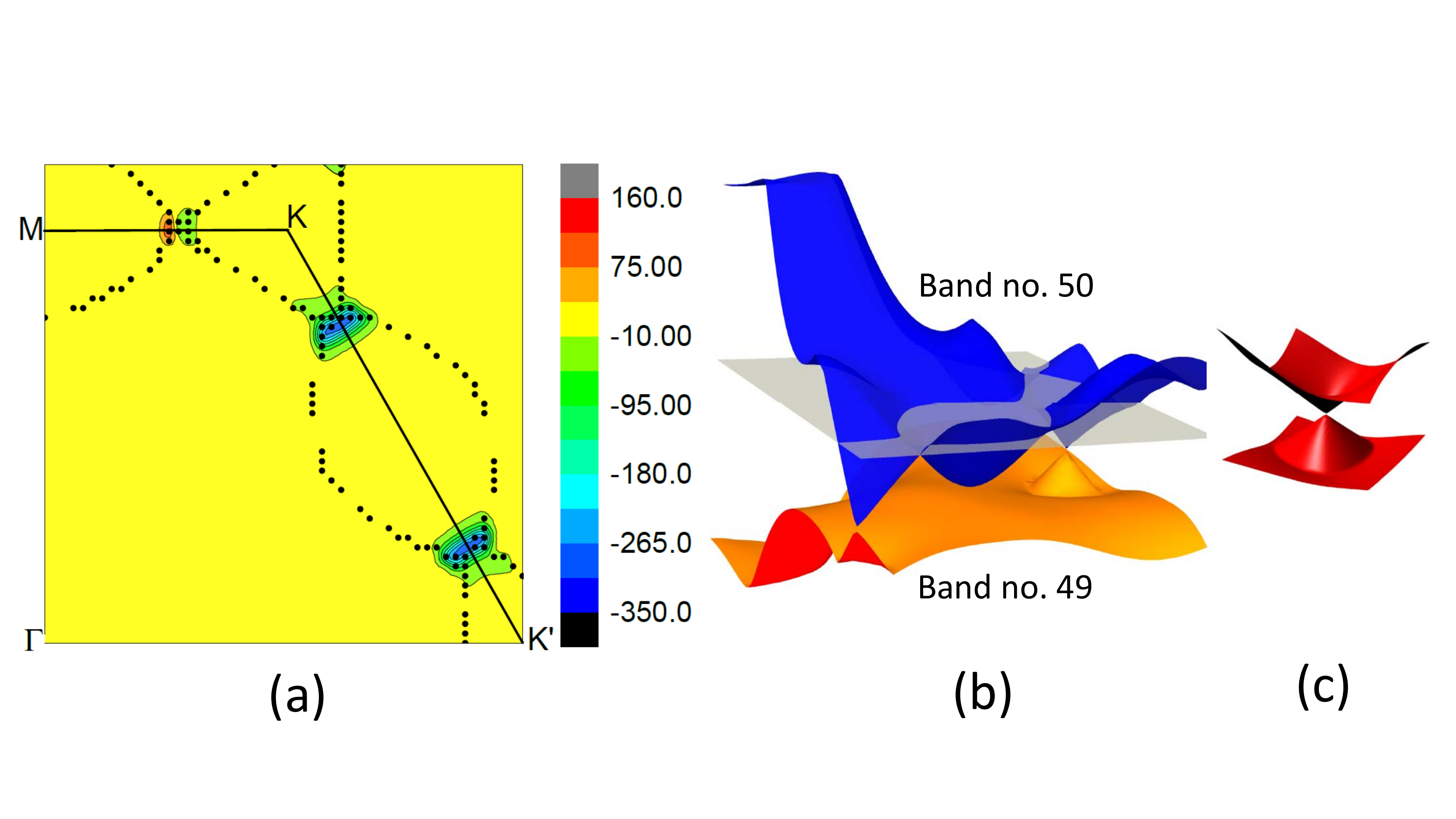}
\caption{\label{fig3} Mn$_3$Sn: (a) Fermi surface contours in the $k_z$ = 0 - plane from band no. 50 and the Berry curvature $\Omega_x(k_x,k_y,0)/2\pi$, special points of the Brillouin zone are marked. The dark-green contours are due to large values of the Berry curvature. (b) Bands no. 49 and 50 near the Fermi energy (gray) in a two dimensional plot. (c) enlarged Weyl point at the point K.  }
\end{center}
\end{figure}
\begin{figure}[h]
\begin{center}
 \includegraphics[width=1.05\linewidth]{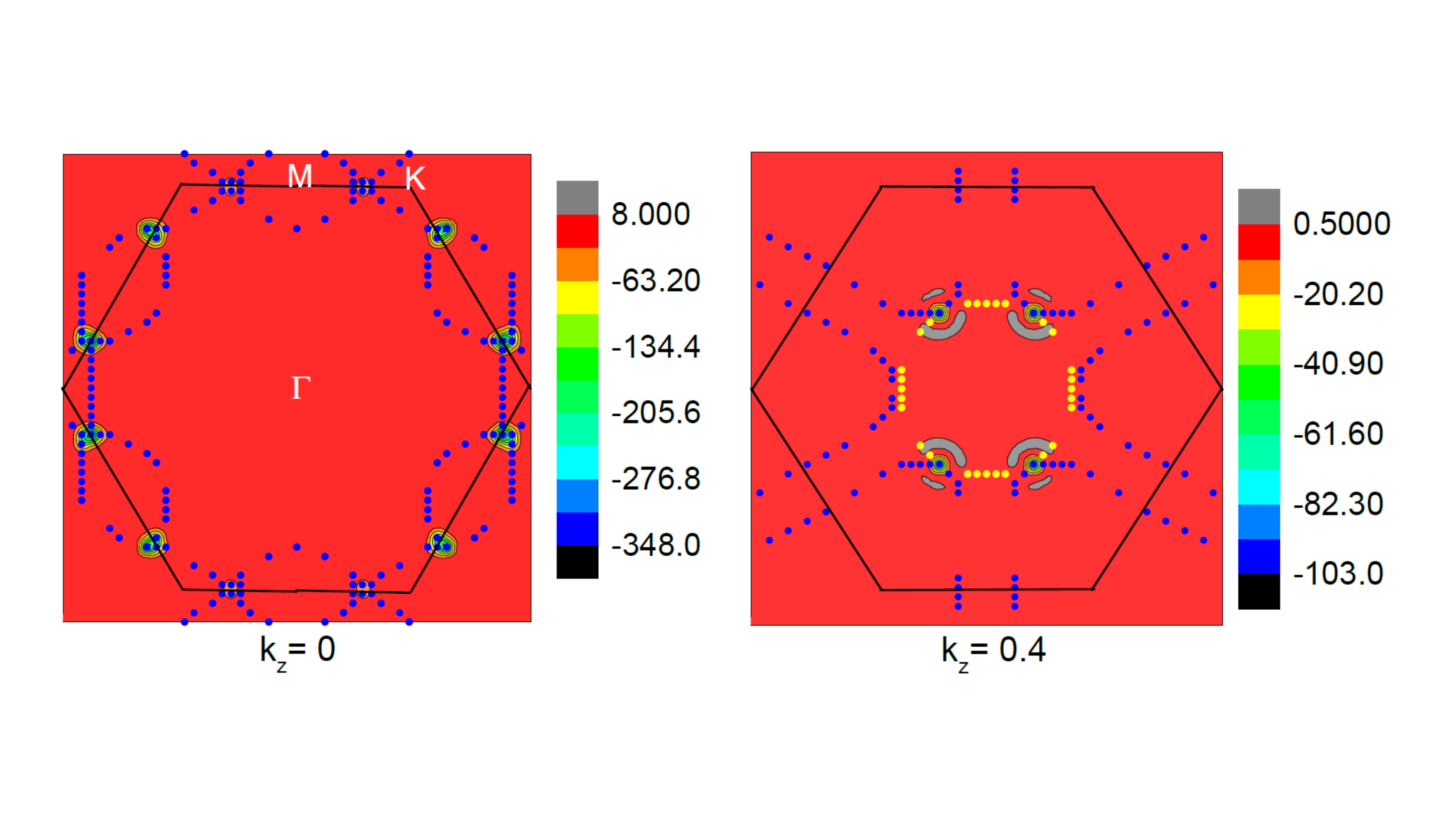}
\caption{\label{fig3-4}Mn$_3$Sn: View of the Fermi surface contours and the Berry curvatures in the extended zone for two planes; as Fig.~\ref{fig3}a $\Omega_x(k_x,k_y,0)/2\pi$ in the $k_z=0$ - plane and $\Omega_x(k_x,k_y,0.4)/2\pi$ in the $k_z=0.4$ - plane. In the latter band no. 51 leads to the Fermi contours plotted in yellow.  }
\end{center}
\end{figure}
\begin{figure}[h]
\begin{center}
 \includegraphics[width=1.0\linewidth]{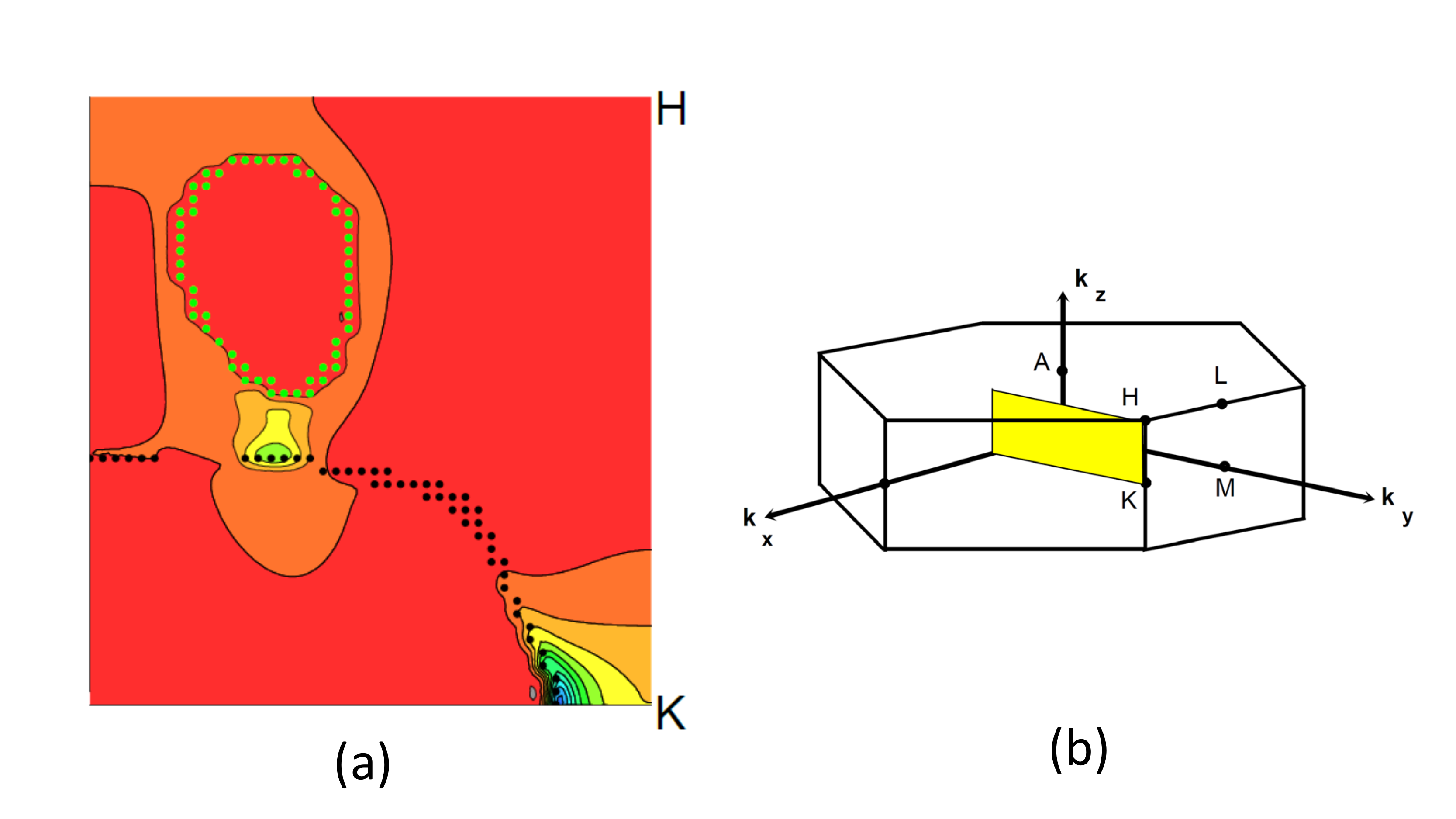}
\caption{\label{fig4}Mn$_3$Sn: (a) Fermi surface contours and the Berry curvature $\Omega_x(k_x=\frac{1}{3},k_y,k_{z})/2\pi$ in the $k_x= 1/3$ - plane. The green contours are due to large values of the Berry curvature. The circular contour graphed with green dots originate from bands no. 51, black dots from band no. 50. (b) Schematic plot of the Brillouin zone showing the plane (yellow) used in (a). }
\end{center}
\end{figure}
\begin{figure}[h!]
\begin{center}
 \includegraphics[width=1.0\linewidth]{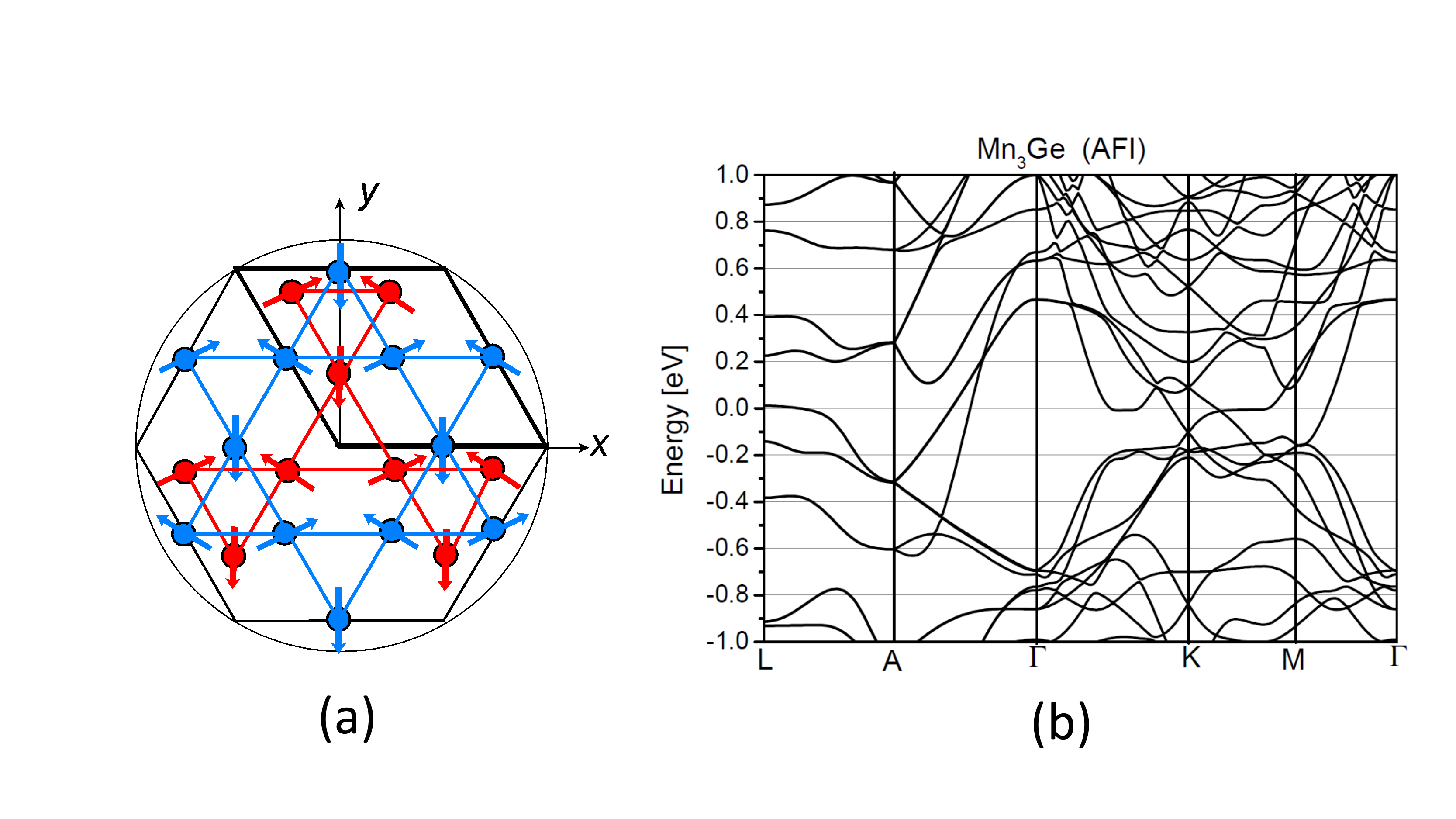}
\caption{\label{fig5} The case of Mn$_3$Ge. (a) Top view of the crystal structure and the directions of the magnetic moments as in Fig. \ref{fig1}. (b) Band structure of Mn$_3$Ge along symmetry lines.}
\end{center}
\end{figure}
\begin{figure}[h!]
\begin{center}
 \includegraphics[width=1.0\linewidth]{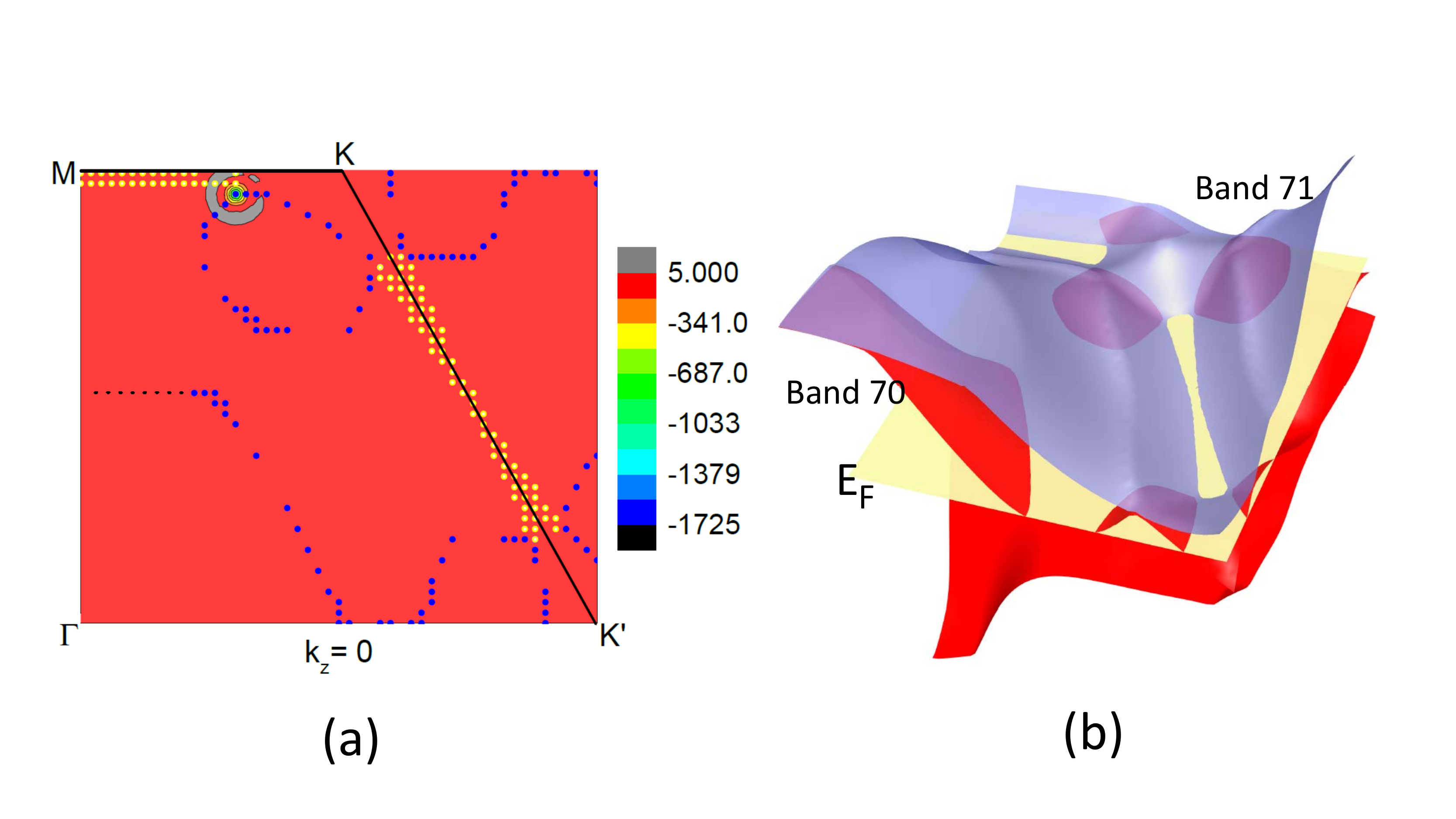}
\caption{\label{fig6}Mn$_3$Ge: (a) Fermi surface contours from band no. 70 (blue dots) and band no. 71 (yellow dots) and the berry curvature $\Omega_y(k_x,k_y,k_{z}=0)/2\pi$ in the $k_z =0$ -plane. Symmetry points of the Brillouin zone are marked. The large value of the Berry curvature shows up as green and black contour. (b) Bands no. 69 to 71 as well as the Fermi energy (yellow) in a two-dimensional graph. }
\end{center}
\end{figure}
\begin{figure}[h!]
\begin{center} 
\includegraphics[width=0.9\linewidth]{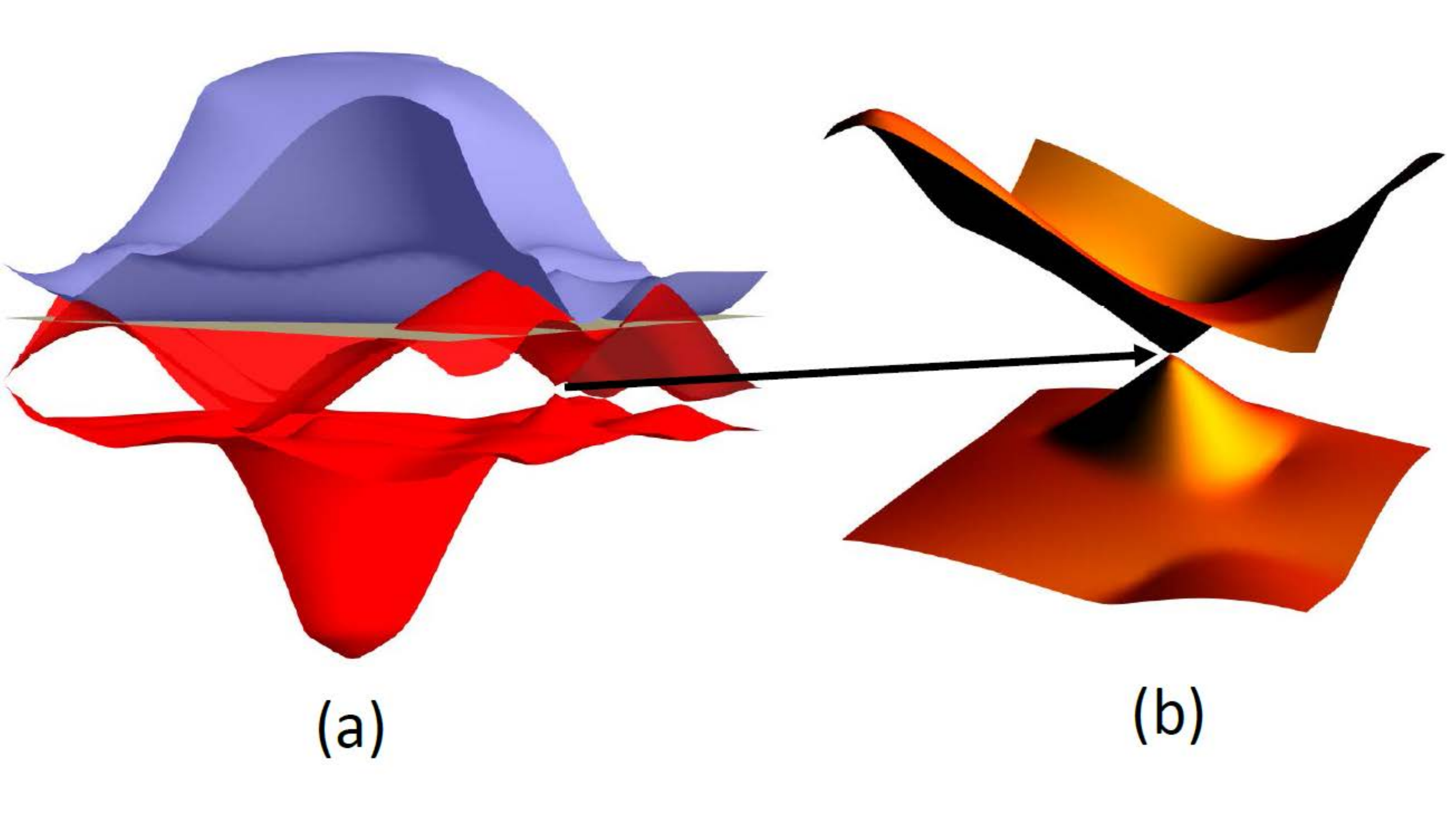}
\caption{\label{fig7}Mn$_3$Ge: (a) Two dimensional bands from band no. 69 to 71 (blue) and the Fermi energy (grey).  (b) The Weyl point at the K point. }
\end{center}
\end{figure}
\begin{figure}[h]
\begin{center}
 \includegraphics[width=1.0\linewidth]{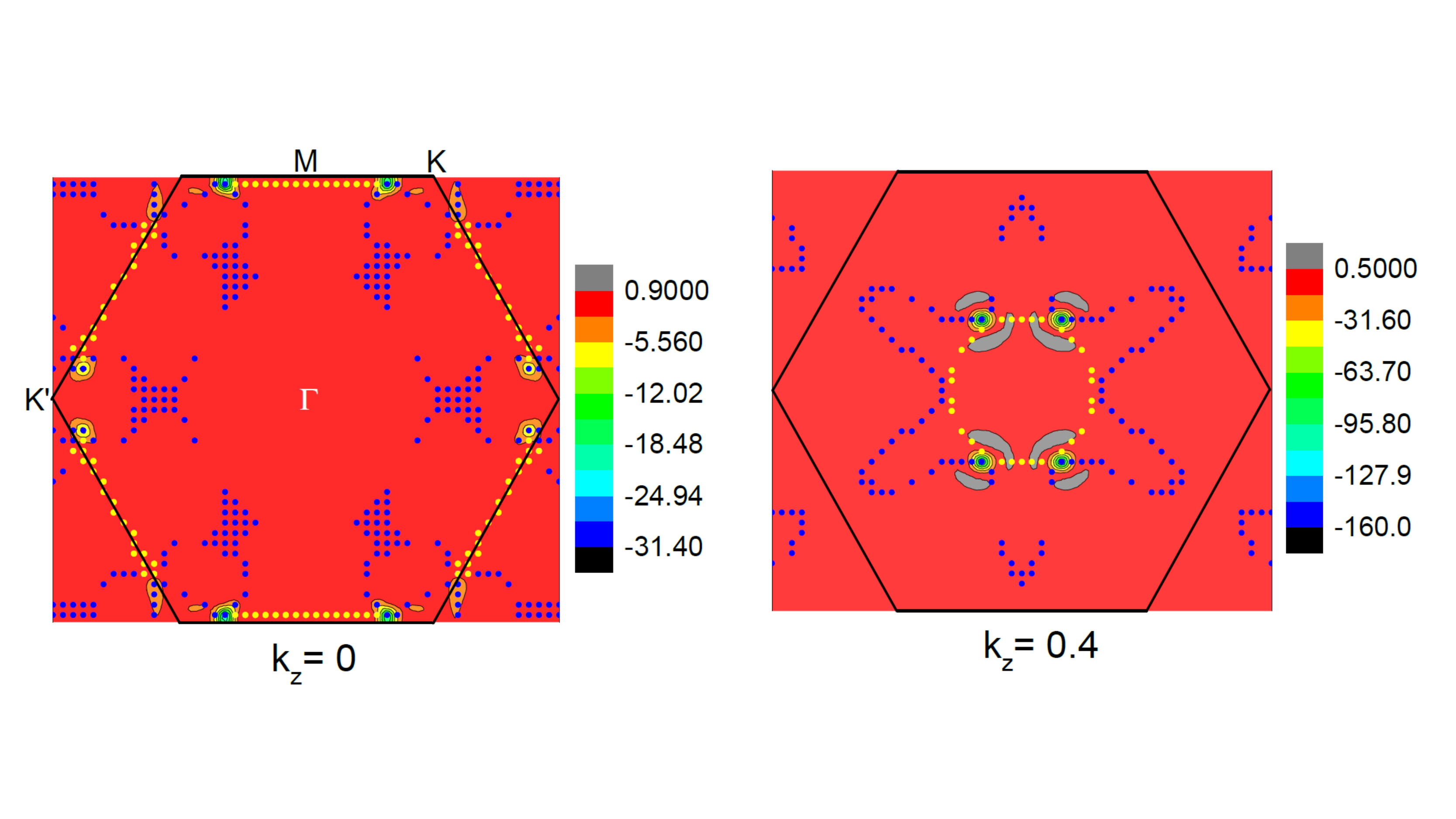}
\caption{\label{fig7-8} Mn$_3$Ge: View of the Fermi surface contours and the Berry curvatures in the extended zone for two planes; as Fig.~\ref{fig6}a $\Omega_y(k_x,k_y,k_{z}=0)/2\pi$ in the $k_z=0$ - plane and $\Omega_y(k_x,k_y,k_{z}=0.4)/2\pi$ in the $k_z=0.4$ - plane. Band no. 71 leads to the Fermi contours plotted in yellow. }
\end{center}
\end{figure}
\begin{figure}[h!]
\begin{center}
 \includegraphics[width=1.0\linewidth]{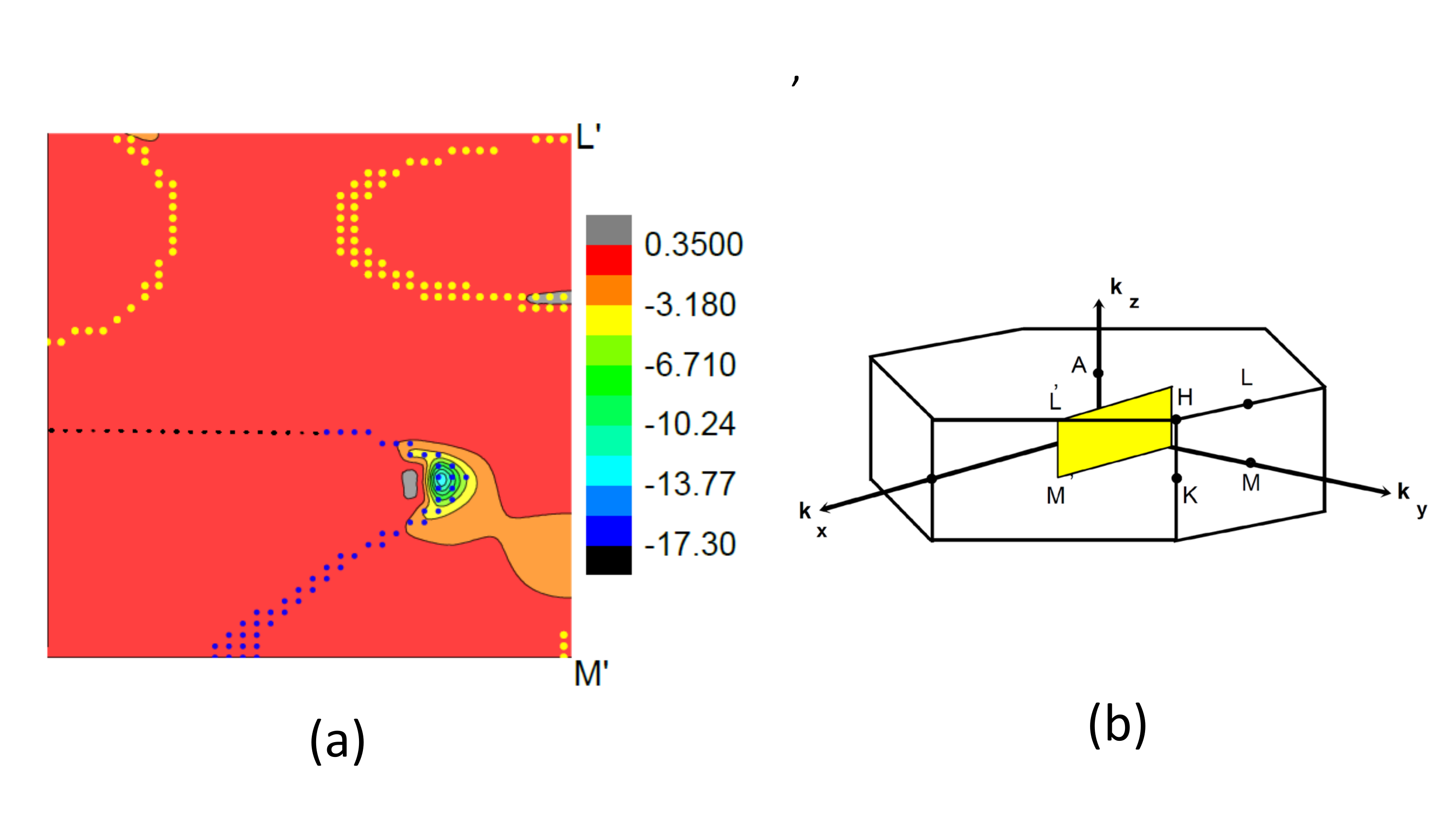}
\caption{\label{fig8}Mn$_3$Ge: (a) Fermi surface contours from band no. 70 (blue dots) and 71 (yellow dots) and the Berry curvature $\Omega_y(k_x,k_y= 1/2\sqrt{3},k_{z})/2\pi$ in the $k_y = \frac{1}{2 \sqrt{3}}$ -plane. Large values of the Berry curvature show up as green contours. (b) Schematic plot of the Brillouin zone showing the surface (yellow) of the plane used in (a).}
\end{center}
\end{figure}

We present results of improved calculations of the anomalous Hall conductivity (AHC), changing the previously used outdated hexagonal coordinate system \cite{KandF}, so that we are now conform with ref.\cite{ajaya} and with the International Tables of Crystallography. Special attention is given to the electronic structure at general points in the Brillouin zone, in order to understand the physical origin of the large AHC. Here we show that topological objects, called Weyl points, might supply the answer. These points are singularities in momentum space, originating from band-crossings. They are understood as magnetic monopoles that are sources of Berry curvature. They come in pairs and give rise to characteristic surface properties, so-called Fermi arcs, which have been predicted theoretically and verified experimentally \cite{solu,wan,turner,yang,liu}.
The ensuing non-trivial topology has been studied in semimetals, but recently the topological approach has  been extended to $\alpha$-iron \cite{vander} and  to the ferromagnetic Heusler compound Co$_2$MnAl \cite{KF2}, which possesses a truly gigantic AHC. Recently, Mn$_3$Sn and Mn$_3$Ge were also seen to host Weyl points that were all found to be below the Fermi energy \cite{binghai}. However, the electrical transport properties and especially the thermoelectric properties studied recently \cite{behnia,arita} are quasiparticle properties  that reside at the Fermi surface as stressed by  Haldane \cite{haldane}. The purpose of this note is to show under which conditions the effect of low-lying Weyl points might appear at the Fermi surface.
Ref. \cite{binghai} contains a thorough study of the surface arcs of Mn$_3$Sn and Mn$_3$Ge but no attempt was made to relate the Weyl nodes to the bulk Fermi surface. Instead of using Haldane's mathematical theory\cite{haldane}, we here present descriptive arguments to show that the Weyl nodes cause the large anomalous Hall conductivity and the distinct thermoelectric effects through the non-zero Chern number of the bands at the Fermi surface. When we say \textit{large} we have in mind values comparable to those of ferromagnetic $\alpha$-iron, \textit{i.e.} of order of 700 S/cm, which is almost achieved in Mn$_3$Ge. 

The anomalous Hall effect is obtained by computing the Berry curvature (BC) in momentum space.  The quantities responsible for the thermoelectric effects are related to the BC as well \cite{haldane}. This is a vector, writing its $p$-component as $\Omega_p(\mathbf{k})$, and is obtained from the curl of the Berry connection,  given by 
$\mathcal{A}(\mathbf{k})=-i\sum_{n \in occ}\ \langle{u_{n{\bf k}}}|\nabla_k|{u_{n{\bf k}}}\rangle$,
where $u_{{n\bf k}}(\mathbf{r})$ is the crystal-periodic
eigenfunction having wave vector \textbf{k} and band index $n$, $\vec{\Omega}(\bf k)=\nabla \times\mathcal{A(\bf k)}$.
The sum extends over the occupied states. 
For its evaluation we use the wave functions from density functional calculations \cite{williams} following ref.\cite{kubler12}, where the numerical work is based on $\Im \ln\det[\langle u_{n \bf k}|u_{m \bf k'}\rangle]$, which is directly related to the Berry connection.\cite{fukui} Spin-orbit coupling (SOC) is treated in second variation and is an essential ingredient. The anomalous Hall conductivity (AHC) follows from the BC by means of
\begin{equation}\label{eq1}
\sigma_{\ell m}=\frac{e^2}{\hbar}
\,\int_{\rm{BZ}}\frac{d\mathbf{k}}{(2\pi)^3}
\Omega_p(\mathbf{k})f(\mathbf{k}),
\end{equation}
where $f(\mathbf{k})$ is the Fermi distribution function, the integral extends over the Brillouin zone,
$\Omega_p(\mathbf{k})$ is the $p$-component of the BC
for the wave-vector \textbf{k} and the components $\ell, m ,p$ (for $x,y,z$) are to be chosen cyclic\cite{xiao}. While this equation is commonly used for the computation of the
AHC, it does not reveal that the transport takes place at the Fermi surface. Haldane \cite{haldane}, as said before, derived another version of eq. (\ref{eq1})
that does just that, but will not be used here. We also need the definition of the Chern number \cite{haldane},
which is given by \begin{equation}\label{eq2}
\frac{1}{2\pi}\int_{\rm BZ}d\mathbf{k}
\Omega^n_p(\mathbf{k})=C_n G^n_p ~. 
\end{equation}
The extra index $n$ appearing here is the state number, which appears in the BC if the sum over  states is omitted. We state that the BZ-integral devided by $2\pi$ gives the Chern number of band $n$, $C_n$, times $G^n_p$, the $p$-component of  a reciprocal lattice vector. The topology of a crystal is non-trivial if  the integer $C_n \ne 0 $. 

\textbf{Mn$_3$Sn} -  We begin with Mn$_3$Sn and show the band structure along symmetry lines in Fig.~\ref{fig2}. It is in good agreement with the band structure shown in Ref.~\cite{claudi}, where the symmetry conditions are given for the anomalöous Hall conductivity and the Süpin Hall conductivity. The integrated Berry curvature vector for the order shown in Fig.~\ref{fig1} a is in the $x$-direction and the calculated AHC  is $\sigma_{yz}=310$ S/cm. The experimental value is obtained at 100 K and is $\sigma_{yz}=100$ S/cm \cite{nakatsuji}. The value calculated in Ref.~\cite{claudi} is $\sigma_{zx}=133$ S/cm obtained for the magnetic order sketched in Fig.~\ref{fig5}a. The difference is not due to the other magnetic order, but is accounted for by our numerical method.  The BC together with contours of the Fermi surface are displayed in Fig.~\ref{fig3} a in the $k_z=0$ -plane. The BC is very large where the green spots
appear. To appreciate the size of the local BC one should multiply the values given in the figure by about 1100 to  convert to units S/cm of the Hall conductivity. The dominating values appear at locations of the Fermi surface where the curvature of the Fermi contours shown by the blue dots is large. Fig.~\ref{fig3} b is a two-dimensional rendering of the band structure and band no. 50 is seen to cut the Fermi energy (gray) thus defining the Fermi contours. The lower band no. 49  touches the higher band in two points in the Brillouin zone. Zooming in at one of the points we see  the Weyl point that is graphed in part (c) of the figure.The band crossing  is at the point K in the Brillouin zone where no signal of the Berry curvature is seen in Fig.~\ref{fig3} a.
Another Weyl point  seen in Fig.~\ref{fig3} b is located at ${\bf{k}}=(2/3,0,0)$. There are all together 6 Weyl points of chirality $\pm 1$ in the extended zone, which is displayed in Fig.~\ref{fig3-4}. We did not try to verify more points described in Ref.\cite{binghai} because they are too far below the Fermi energy.
In Fig.~\ref{fig3-4} we look at the  Fermi contours in the extended zone to visualize the development of the BC in the $k_z=0.4$ - plane from the BC in the $k_z=0$ - plane. In the $k_z=0.4$ - plane a second contour appears (yellow dots) due to band no. 51. The contours due to band no. 50 (blue dots) deform and carry the BC spots at the location of large curvature of the Fermi contours.  
In Fig.~\ref{fig4}(a) the Berry curvature together with the Fermi contours is displayed in the $k_x=1/3 $ - plane which is sketched schematically in part
(b) of the figure. The strong signal at the Fermi energy is a reflection of the Weyl point at K.  The Fermi contour that shows up in the upper part of Fig.~\ref{fig4} a is due to band no. 51. There is no BC emanating from this contour.

\textbf{Mn$_3$Ge} - We continue with Mn$_3$Ge. The magnetic structure, shown in Fig.~\ref{fig5} a, is different from that of Mn$_3$Sn. One is obtained from the other by rotating all spins by $90^0$ about the $z$-axis. Computationally the total energies of the two structures in either one of the systems are very close so that we  rely on experimental information for the present assignment.
The integrated Berry curvature vector in this case is in the $y$-direction and the calculated AHC amounts to $\sigma_{zx}=560$ S/cm. This is to be compared with low-temperature values of $\sigma_{zx}=500$ S/cm measured by both groups~\cite{ajaya,naoki}. The calculated value given in Ref.\cite{claudi} is $\sigma_{zx}=330$ S/cm. The band structure along symmetry lines is given in Fig.~\ref{fig5} b and is in good agreement with Ref.~\cite{claudi}. It is nearly identical to Fig.~\ref{fig2}, except for minute differences near the Fermi energy at K and M. These matter as can be seen in Fig.~\ref{fig6} a, where Fermi contours are displayed together with the BC in the $k_z=0$ -plane. Now the Fermi surface is formed by two bands, \textit{viz.} bands nos. 70 and 71. A spot of large value of the BC appears as green and black contours near a Fermi contour that curve strongly below the line M K. Fig.~\ref{fig6} b gives the band structure near the Fermi energy in a two dimensional plot. In Fig.~\ref{fig7} a we locate a band crossing between band nos. 69 and 70. Zooming in near K  we see the Weyl
point shown in Fig.~\ref{fig7} b. Another Weyl point appears at K' = (2/3,0,0).
As for Mn$_3$Sn  there are 6 Weyl points of chirality $\pm 1$ in the extended zone. The other Weyl points described in Ref.\cite{binghai} we ignored since they are too far below the Fermi energy. In Fig.~\ref{fig7-8}
 as before we look at the  Fermi contours in the extended zone to visualize the development of the BC in the $k_z=0.4$ - plane from the BC in the $k_z=0$ - plane.  The contours due to band no. 70 (blue dots) deform and carry the BC spots at the location of large curvature of the Fermi contours, whereas the contours due to band no. 71 contract to the yellow circle.
Fig.~\ref{fig8} a, finally, shows Fermi contours together with the BC in the $k_y = \frac{1}{2 \sqrt{3}}$ -plane, which is shown in part (b) of the figure. The green BC spot in (a) is a reflections of the  Weyl points at K and K'. The yellow Fermi surface contours near L' show no sizeable BC spots. 
   
\textbf{Discussion} - 
We now try to explain the spots of large BC, for short \textit{'hot spots'}, with the physical concepts developed for Weyl semimetals\cite{turner}. It is not \textit{a priori} clear if these concepts also apply to noncollinear antiferromagnetic metals, but we assume they do. There is first the fact that the Chern number $C$  of a band is non-zero between two Weyl points of opposite chirality. Thus the band no. 50 for Mn$_3$Sn (see Fig.~\ref{fig3}) and band no. 70 for Mn$_3$Ge (see Fig.~\ref{fig6} and \ref{fig7}) at the Fermi energy have $C\ne 0$. That is \textit{the hot spots are a property of the non-zero Chern number}. The next higher band, no. 51 for Mn$_3$Sn and no. 71 for Mn$_3$Ge,  do produce  uniform BC but no spots. A simple formula valid for a model Weyl semimetal relates the AHC to the difference vector between two Weyl points of opposite chirality \cite{turner}. Using the Weyl points obtained here, we obtain about $\sigma_{yz}= 230$ S/cm for Mn$_3$Sn and $\sigma_{zx}=390$ S/cm for Mn$_3$Ge. It must be remarked that these values are fortuitously close to the observed ones, since none of the assumptions underlying  their derivation are valid here, except for the Chern number of the bands. But the range of validity may be larger than assumed. This opens another question: Could it be that the chiral anomaly exists for noncollinear antiferromagnets? Presumably this question can only be answered by experiments. Concerning  the thermoelectric effects it is to be seen if they are as large in Mn$_3$Ge as in Mn$_3$S.

\end{document}